# Associated production for the Standard Model Higgs at CDF

**Daniela Bortoletto**

Physics Departments, Purdue University, West Lafayette, IN 47906, USA

bortolet@purdue.edu

**Abstract**. We report the latest result for the search for the standard model higgs produced in association with a W and a Z boson at CDF. The results include about 1 to 1.7 fb-1 of data collected by CDF in run II of the Tevatron. Novel analysis techniques have been developed to enhance the sensitivity of these searches.

## 1. Introduction

The Higgs boson is an essential component of the Standard Model (SM) which explains the origins of mass and electro-weak symmetry breaking. Precision measurements [1] indicate that the mass of the Higgs boson should be less than 144 GeV/c$^2$ and is therefore within the reach of the Tevatron if enough luminosity is collected. Previous searches for the Standard Model Higgs set limits on its production cross section at $\sqrt{s} = 1.8$ and $\sqrt{s} = 1.96$ GeV/c [2]. Here we report the most recent results on the search for the standard model Higgs produced in association with a *Z* or a *W* obtained by the CDF collaboration with 1 to 1.7 fb$^{-1}$ of data. This search focuses on a Higgs with mass below 135 GeV/c$^2$ which decays mainly into $b\bar{b}$. The search is divided into three optimized analysis. The first considers decays $Z \rightarrow \ell\ell$ and therefore requires final states with two leptons. The second requires a lepton from the W decay and transverse missing energy ($\slashed{E}_T$). The last one considers $Z \rightarrow \nu\bar{\nu}$ or $W \rightarrow \ell\nu$ when the lepton was undetected leading to a final state with large $\slashed{E}_T$.

All searches require at least two hadronic jets and that one of the two jets must be consistent with the hadronization of a *b*-quark. Two algorithms are used at CDF for *b*-jet identification. The SECVTX algorithm identifies a *b*-jet by finding secondary vertices inside a jet. It can be used with two different settings called tight and loose according to their identification efficiency and purity [3]. The second algorithm, JETPROB, identifies b-jets by requiring that the tracks in the jet have a low probability of originating from the primary vertex [4].

## 2. The dilepton channel

The search for the Higgs boson in $ZH \rightarrow \ell\ell b\bar{b}$ with 1 fb$^{-1}$ of data uses both $Z \rightarrow e^+e^-$ and $Z \rightarrow \mu^+\mu^-$. The search in this channel [5] is characterized by a small background due to the requirement of two leptons and the Z constraint.

CDF has improved considerably the reach of this search by loosening the lepton identification requirements with respect to the one used by other high physics analysis. We require two jets with

$|\eta|< 2$ and $E_T > 15 GeV$. We select events with two loose *b*-tags or one tight *b*-tag. Since the only source of $\not{E}_T$ in $ZH \to \ell\ell b\bar{b}$ is coming from mismeasurements of jet energies, the two candidate Higgs jets are corrected with an Neural Network optimized to reassigns the $\not{E}_T$ to the jets. This correction is derived from *ZH* Monte Carlo and improves the jet energy resolution from 16% to 10%.

Backgrounds to this search such as $Z+Jets$, $t\bar{t}$ and diboson production were evaluated using Monte Carlo. Backgrounds due to misidentified leptons and misidentified *b*-jets were evaluated using data. To maintain efficiency and improve signal (S) over background (B) we employ a 2-D neural network trained to discriminate *ZH* from $Z+Jets$ on one axis and *ZH* from top on the second axis. Training is done on a range of Higgs masses from 60 to 100 $Gev/c^2$.

Several sources of systematic uncertainty on the signal and the background were evaluated. They include uncertainties due to the normalization of $Z+c\bar{c}$ and $Z+b\bar{b}$ (40%), the cross section of $t\bar{t}$, *WZ* and *ZZ* (20%), the mistag background (8%), the b-tagging efficiency (8%), the trigger efficiency (1%), the misidentified leptons (50%), the integrated luminosity (6%), the jet-energy scale, and kinematic distribution for the $Z+Jets$ background predicted by different generators.

For 2 loose tags (1 tight tag) we expect 7.36±1.56 (60.8±10.7) background events and we observed 6 (54) where we expect 0.13 (0.25) ZH events for a Higgs mass of 115 GeV/$c^2$. No evidence is found for Higgs production above the SM backgrounds and limits on the Higgs production are set for several Higgs masses as shown in Fig 1 (a).

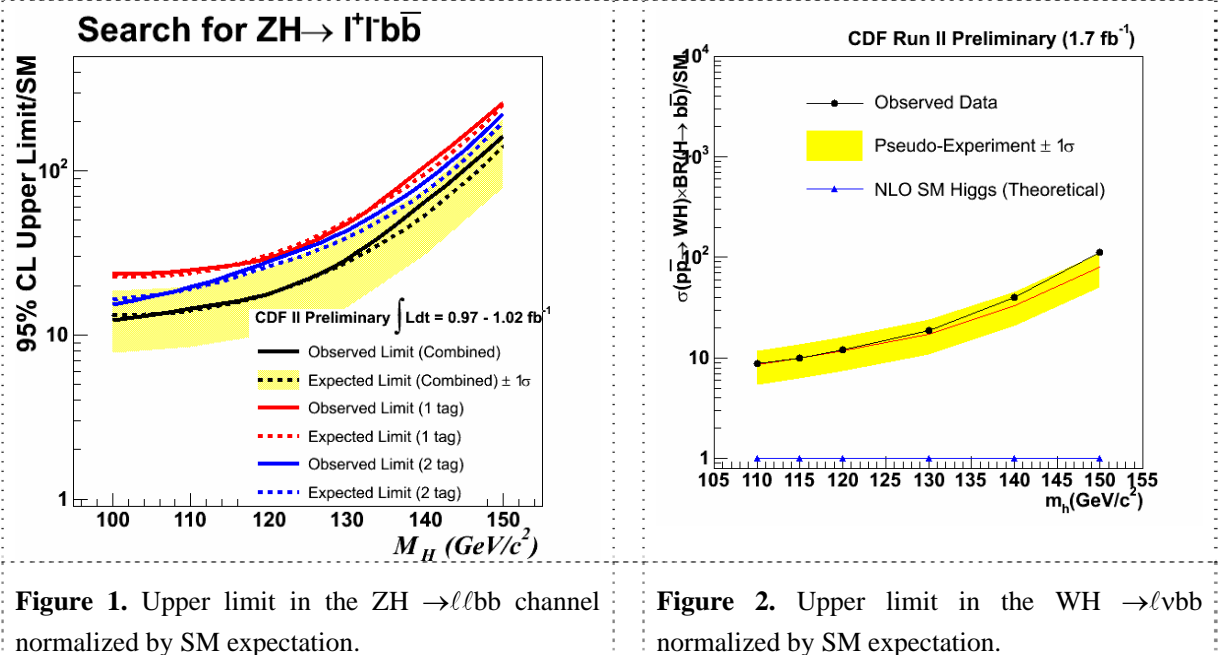

**Figure 1.** Upper limit in the ZH →$\ell\ell$bb channel normalized by SM expectation.

**Figure 2.** Upper limit in the WH →$\ell$νbb normalized by SM expectation.

### 3. The lepton and missing energy channel

The search for the Higgs boson produced in association with a W boson is performed on 1.7fb[-1][6]. We select events consistent with $W \to \ell\nu$ by requiring an identified lepton with $E_T$> 20 GeV or $p_T$>20 GeV/c and large missing transverse energy ($\not{E}_T$ >20 GeV). The search has been optimized to increase $H \to b\bar{b}$ sensitivity by requiring two SECVTX tight tags or one tight SECVTX tag and one JETPROB tag. These categories are considered separately to take advantage of the different S/N ratios in the two types of double tag events. For a Higgs mass of 110 GeV/$c^2$, the acceptance is (0.43 ±0.05)% and

(0.36 ±0.06)% for a SECVTX double tag and for a SECVTX - JETPROB double tag respectively. Since the two categories are implemented to be mutually exclusive, the total acceptance is given by the sum of the two double tag types.

The background to this search is dominated by falsely *b*-tagged events, *W* production with heavy flavor quark pairs, QCD with misidentified *W*, top quark production and electro-weak production of dibosons and single top. The backgrounds are determined using a combination of Monte Carlo and data based techniques. The agreement found in estimating the background level in the *W+1, 3,* and *4 jets* confirms our background estimate in the *W+2 jets* sample which is used for our search.

Systematic uncertainties considered in this search are due to the signal acceptance because of the modeling of initial and final state radiation (5%), b-tagging (9%), jet energy scale (3%), parton distribution functions (2%), trigger efficiency (<1%), and lepton identification (2%).

An artificial neural network which includes the dijet invariant mass, the total system $p_T$ and the event $p_T$ imbalance is used to improve the discrimination between the Higgs signal and the large backgrounds in the *W* + 2 *jets* bin. Since the SM background without the Higgs accounts for the observed number of events, we set limits on the Higgs production separately for events in the two double *b*-tags categories. The limits are then combined to give the final limit shown in Fig. 2 as the function of the Higgs mass. The neural network improves the limit by approximately 10%, yielding $\sigma(p\bar{p} \to WH) \times BR(H \to b\bar{b}) < 1.4$ to 1.3 pb for Higgs masses from 110 GeV to 150 GeV.

## 4. The missing energy channel

The search for the Higgs in the missing energy channel uses $0.97 \pm 0.06$ fb$^{-1}$ of data [7]. We consider a scenario where the Higgs is produced in association with a *Z* or *W* leaving only $\not{E}_T$ in the detector. We require the events to have $\not{E}_T > 50$ GeV and two jets with $E_T > 20$ GeV and pseudorapidity $|\eta|$ <2.4. The azimuthal separation of the two jets is required to be larger 1.0 radian. Moreover one of the two jets must be central with $|\eta|< 0.9$ and the jet of highest (second highest) energy in the event must have $E_{T,1}>35$ GeV ($E_{T,2}>25$ GeV). We also require one tight *b*-tag or two loose *b*-tags

In order to avoid potential bias in the search, we test the SM background in control regions. The backgrounds are multi-jets, top, electroweak bosons (*W* and *Z*) with additional partons, and electroweak dibosons. Control region 1 is dominated by the multi-jet background. The events in this region do not contain any identified leptons and the azimuthal angular separation $\phi$ between the second highest jet and the $\not{E}_T$ is less than 0.4. Control region 2 is selected by requiring at least one lepton and $\phi (E_{T,2}, \not{E}_T )>0.4$. This region is sensitive to electroweak and top decays. The extended signal region contains events have no identified high-p$_T$ lepton, $\phi (E_{T,2}, \not{E}_T )>0.4$ and $\phi (E_{T1}, \not{E}_T )>0.4$. We use control region 1 to normalize the QCD heavy flavor background generated with PYTHIA to the data. The light flavor contribution is estimated using a mistag matrix determined in data. Control region 2 is sensitive to the modeling of electroweak backgrounds. Distributions of kinematic variables have been studied and found to be in agreement with observations [3] in the control regions.

We optimize the Higgs sensitivity by requiring $\phi (E_{T1}, \not{E}_T )>0.8$, $\not{H}_T / H_T >0.45$, $E_{T1} >60$ GeV, and $\not{E}_T >70$ GeV. The variables $H_T$ and $\not{H}_T$ are the scalar sum and the magnitude of the vectorial sum of the two jets. For a Higgs boson of 115 GeV/c$^2$ we expect a total of 1.2 (0.4) Higgs events with one (two) *b*-tagged jets. We observe 268 exclusive single *b*-tagged events, which is in agreement with SM background expectations of 251±43 events. Requiring at least two *b*-tagged jets we observe 16 events, where 14.8±2.7 are expected.

The systematic uncertainty on the signal and the background predictions is studied in detail in [7]. The dominant correlated uncertainties affecting both background and signal predictions is the jet energy scale which varies between 10% and 20 % for multi-jets, 15% for dibosons, 26% for *W*+heavy flavor, 17% for *Z*+heavy flavor, 1% for top, and 8% for the signal events. The dominant uncorrelated systematic uncertainties are due to the cross sections and the Monte Carlo statistics. We assign 11% to the top, 11.5% to the diboson and 40% to the $W$ and $Z$ + jets cross sections. Uncertainties due to limited Monte Carlo statistics are 32% on the multi-jet, 20% on the $W$, and 11% on the $Z$ + jets predictions.

Since no significant excess is observed, we compute 95% C.L. observed and expected limits for the Higgs cross section. The limits are computed separately for single and double tagged events and for WH and ZH, and then are combined by taking the product of their likelihoods and varying simultaneously the correlated uncertainties. The limits are shown in fig. 3

## 5. Conclusions

The CDF collaboration is continuing to improve its sensitivity for the Higgs boson by updating analysis tools and techniques. The results from the analyses presented here and similarstudies performed by the D0 experiments was combined [8] as shown in fig. 4.

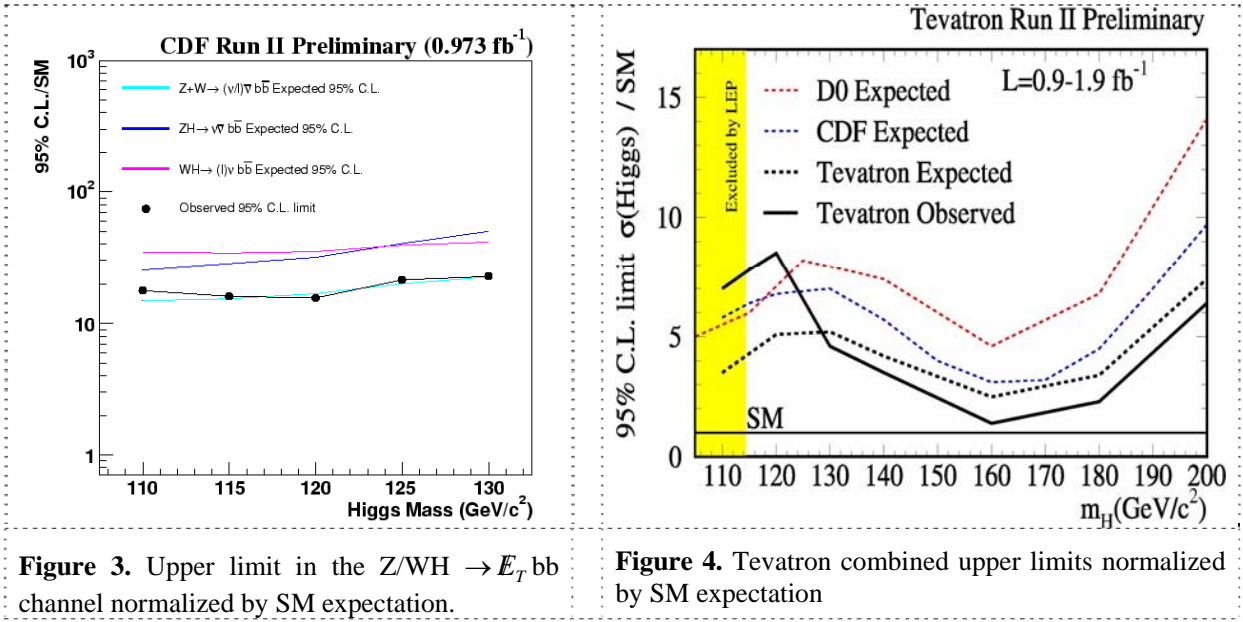

**Figure 3.** Upper limit in the Z/WH $\to \slashed{E}_T$ bb channel normalized by SM expectation.

**Figure 4.** Tevatron combined upper limits normalized by SM expectation